# Selective Adsorption and Chiral Amplification of Amino Acids in Vermiculite Clay -Implications for the origin of biochirality.

Donald G. Fraser*[1], Daniel Fitz[2], T. Jakschitz[2] and Bernd M. Rode[2,]

[1]Department of Earth Sciences, University of Oxford, Parks Road, Oxford OX1 3PR, United Kingdom
[2]Faculty of Chemistry and Pharmacy, Institute of General, Inorganic and Theoretical Chemistry,
University of Innsbruck,
Innrain 52a, A-6020 Innsbruck, Austria

* Corresponding author
Email:  don@earth.ox.ac.uk

**Abstract**

Smectite clays are hydrated layer silicates that, like micas, occur naturally in abundance. Importantly, they have readily modifiable interlayer spaces that provide excellent sites for nanochemistry. Vermiculite is one such smectite clay and in the presence of small chain-length alkyl-$NH_3Cl$ ions, forms sensitive, 1-D ordered model clay systems with expandable nano-pore inter-layer regions. These inter-layers readily adsorb organic molecules. N-propyl $NH_3Cl$ vermiculite clay gels were used to determine the adsorption of alanine, lysine and histidine by chiral HPLC. The results show that during reaction with fresh vermiculite interlayers, significant chiral enrichment of either L- and D-enantiomers occurs depending on the amino acid. Chiral enrichment of the supernatant solutions is up to about 1% per pass. In contrast, addition to clay interlayers already reacted with amino acid solutions resulted in little or no change in D/L ratio during the time of the experiment. Adsorption of small amounts of amphiphilic organic molecules in clay inter-layers is known to produce Layer-by-Layer or Langmuir-Blodgett films. Moreover atomistic simulations show that self-organization of organic species in clay interlayers is important. These non-centrosymmetric, chirally active nanofilms may cause clays to act subsequently as chiral amplifiers, concentrating organic material from dilute solution and having different adsorption energetics for D- and L-enantiomers. The additional role of clays in RNA oligomerization already postulated by Ferris and others, together with the need for the organization of amphiphilic molecules and lipids noted by Szostak and others, suggests that such chiral separation by clays in lagoonal environments at normal biological temperatures might also have played a significant role in the origin of biochirality.

## Introduction

Elucidating the origin of biological homochirality has occupied a central position in studies of the origin of life[1] since the discovery of chiral[2] symmetry by Pasteur[3-4]. Almost all DNA-coded amino acids in proteins and enzymes are L-isomers[5] and the sugars in DNA and RNA are right-handed. Intriguingly, recent work has shown that the two might be related in that L-amino acids can also catalyse the formation of D-glyceraldehyde, and hence D-sugars, in plausible planetary environments[6]. Proposed mechanisms for the origin of biochiral specificity include the templating of organic precursors on the surfaces of non-centrosymmetric minerals such as quartz and calcite [7-9]; chance selection of one isomer followed by sequential replication by the resulting living process[10]; asymmetry at the central Cu ion of the complex involved in the salt-induced peptide formation reaction[11]; differential crystallization rates of D- and L-isomers [12-13, 14-15], seeding from space with non-racemic precursors[16-18] synthesised under conditions involving circularly polarised ultra-violet light [19-20] and even mechanisms involving possible quantum mechanical condensation phenomena at low temperature and the weak nuclear force [21-22].

There is, however, no consensus as to how such near-perfect enantiomeric separation occurred in biology and in the absence of a net chiral symmetry operator, no simple fractionation mechanism can lead to overall chiral selectivity. Mechanisms involving partitioning onto chiral crystallographic substrates can only succeed on their own if there is a preponderance of one form or the other in nature. Straightforward mechanisms involving seeding from space suffer both from the low overall abundances of organic molecules introduced with meteorites and the likelihood of their partial destruction during entry through the Earth's atmosphere. Moreover, in the absence of mechanisms for chiral protection, selection or amplification, any primordial enantiomeric excess would be likely to be racemized quickly [5, 23].

Nevertheless, significant chiral excesses of the amino acid L-isovaline have recently been reconfirmed in samples of the CM-type[24] meteorite Murchison (up to 18.5%) and the CI-type[24] meteorite Orgeuil (up to 15.2%) [25]. Importantly, isovaline is not naturally present in the terrestrial biosphere and the excess of L-isovaline over D-isovaline cannot have been introduced by terrestrial contamination. The relative distribution of the organic material observed in the meteorite samples [26] and in particular the ratio of β-alanine to α-isobutyric acid (AIB)[25], is correlated with the degree of aqueous alteration observed mineralogically. Aqueous alteration occurred in both parent bodies and at least in the case of CM is thought to have been caused by the melting of frozen $H_2O$ inside the parent body caused by heat from short-lived radioactive isotopes like $^{26}$Al and $^{60}$Fe [27] and to have lasted $10^2$ to $10^4$ years some $4.55 \times 10^9$ years ago. Aqueous alteration is also related to the chiral excesses observed so that the β-alanine/AIB ratio correlates with the L/D chiral excess of isovaline. These data thus reaffirm and extend previous work on the correlation of enantiomeric excesses with aqueous alteration and the presence of hydrated phases in meteorite samples [17, 28].

Clay minerals form dynamic hydrous systems with d-spacings that vary with water activity[29-31]. Because of their high surface areas and adsorption characteristics, clays have long been postulated to play a key role in concentrating the products of non-biological organic chemical processes leading to the Huxley-Darwin-Oparin-Haldane model for the origin of living systems [32-33]. Clays can also act as catalysts for the formation of RNA oligomers from activated nucleotide precursors [34-35]. Furthermore, montmorillonite clay and alumina both catalyse polymerisation of amino acids in peptide formation reactions such as those of glycine, alanine, leucine, valine and proline[36-41].

All low temperature wet aluminosilicate planets like Earth contain clay minerals. The correspondence of the catalytic role of clay surfaces in activated RNA nucleotide oligomerization, together with the need for the organization of amphiphilic molecules and lipids noted by Szostak and others and the observation of enantiomeric excesses by aqueous alteration in the presence of hydrated phases in meteorite samples, make experimental and theoretical studies of the possibility of chiral selectivity by clays and other hydrated phases of great interest. The layered structure of clays could also allow them to have played a key role in concentrating, shielding and catalysing the assembly of some of the most important molecules necessary for the development of living systems that can metabolize, reproduce and evolve [42-45].

In the present paper we report the results of amino acid adsorption experiments carried out using a sensitive, 1-dimensionally-ordered vermiculite clay gel. Vermiculite clay is common in nature and because of its adsorption characteristics, is widely used as an adsorbent and in its heated, expanded form, as thermal insulation. It can be ion exchanged by replacing the inorganic cations present in the natural mineral with organic cations such as n-propylammonium ions to produce sensitive 1-dimensionally ordered gels with very large d-spacings. These have well defined (00l) Bragg peaks that are determined by the ordered layers and, in the case of the n-propylammonium vermiculite used here, d-spacings which vary with temperature and solution concentration up to as much as 900Å[46-47]. Because of the weak inter-layer bonding, the d-spacings of these n-propylammonium gels are very sensitive to subtle changes in electronic environment including those caused by chemisorption, making them excellent test-beds for clay-organic nanochemistry. This makes them highly suitable for exploring the ways in which clay minerals interact with chiral organic materials. Recent neutron scattering studies[48] have shown that the interlayer spacings of ordered vermiculite clay gels respond sensitively to the addition of D- and L-histidine solutions and exhibit chiral selectivity. Because of their sensitive response to small changes in bonding environment, the use of these vermiculite clay-gels makes it possible to study processes that would otherwise be very difficult and require considerable time or even be experimentally intractable using other materials. These include the chemical processes that, over time, may have led to the development of metabolizing and self-replicating systems in, for example, hydrous, lagoonal environments.

The present experiments were designed to use the sensitivity of the vermiculite gel system to investigate the adsorption of chiral amino acids, using chiral HPLC to measure any changes and amplification of the D/L ratios of amino acid solutions each reacted with identical vermiculite clay gels.

**Experimental**

Crystalline vermiculite is formed by natural weathering of micas like phlogopite. It is a magnesium aluminosilicate clay mineral with a 2:1 sheet structure of tetrahedrally (T) and octahedrally (O) coordinated polyhedra (TOT) having the general formula $[Si_{4-x}Mg_3O_{10}.(OH)_2]M_{4x}^+.nH_2O$. Cation vacancies and substitutions in the aluminosilicate sheets lead to negative charge that is balanced by hydrated cations in the interlayer sites. The vermiculite used in this study comes from Eucatex in Brazil and occurs naturally as flakes of approximately 10 mm x 10 mm x 2 mm with the formula $Si_{3.07}Mg_{2.72}Al_{0.83}Fe_{0.25}Ti_{0.06}Ca_{0.13}Cr_{0.005}K_{0.005}O_{10}(OH)_2Na_{0.65}.nH_2O$ [49]. These flakes were ion-exchanged for 1 year in 1M NaCl solution to remove the naturally occurring inter-layer counterions[50]. The Na ions were then replaced with n-propyl $NH_3^+$ organic cations by soaking the vermiculite in $C_3H_7 \cdot NH_3Cl$ (Sigma-Aldrich 242543, >99%) solution over a period of several years.

Previous studies have shown that replacement of the inorganic interlayer cations with organic cations such as propylammonium and butylammonium leads to the formation of highly 1-dimensionally ordered gels that can absorb very large quantities of water in the interlayer region. These gels have well-characterised 1-dimensional d-spacings that range up to 900Å and there is a reversible crystal-gel phase transition at temperatures from around 14C to 36C that depends on the concentration of the interlayer ions. Previous work has established the relationship between *d*-spacing and concentration in propylammonium chloride vermiculite gels used in this study[51]. In the region of most interest, around 0.35M, the equilibrium *d*-spacing increases by around 1 Å for a 0.01M decrease in concentration and the d-spacing itself is of order 50Å depending on temperature.

The delicate gels used in the experiments were again exchanged before use for 48 hours in fresh 0.35M $C_3H_7 \cdot NH_3Cl$ (Sigma Aldrich, >99 %) solution at 25°C. Solutions of 50:50 racemic mixtures of 0.05M alanine, lysine and histidine (Bachem, >99 %) in 0.35M $C_3H_7 \cdot NH_3Cl$ were prepared using the same stock 0.35M $C_3H_7 \cdot NH_3Cl$ solution. All water used was ultrapure HPLC purity (ROTISOLV HPLC gradient grade, Carl Roth GmbH, Germany) and nitrile gloves were worn at all stages during handling.

Individual flakes of $C_3H_7 \cdot NH_3Cl$ - vermiculite gel, weighing when wet ~100mg, were selected from solution and placed vertically in clean glass HPLC vials, covered immediately with 1.5ml racemic alanine, lysine or histidine solutions respectively, made up as described in 0.35M $C_3H_7 \cdot NH_3Cl$ and sealed with a crimp top seal. Previous neutron scattering studies showed that diffusion to achieve steady state as measured by change in d-spacing takes about 2 hours under these conditions[48]. Samples were sequentially reacted for 2 hours after which an approximately 50μl sample was taken, filtered through a 0.22μm hydrophilic PVDF syringe filters (Carl Roth GmbH, Germany) and frozen for HPLC analysis in 2ml HPLC vials with low-volume inserts at -30C. The supernatant amino acid solution in contact with the first clay gel was immediately decanted into a fresh HPLC vial containing a fresh piece of vermiculite gel taken from its stock 0.35M $C_3H_7 \cdot NH_3Cl$ solution and the vial was resealed as before for reaction. Temperatures were maintained constant at 32C +/- 0.1C using an Eppendorf Thermomixer comfort dry thermostatic block except when changing gel and taking an aliquot for analysis.

Amino acid analyses were performed by HPLC on an Agilent 1100 series system equipped with a diode array detector. Solvent A was 2mM $CuCl_2$ (Merck, Darmstadt, Germany) in ultrapure water and solvent B was pure acetonitrile (ROTISOLV HPLC gradient grade, Carl Roth GmbH, Germany). 2μl of the samples were injected into a chiral Phenomenex Chirex 3126 column (150×4.6mm) equipped with a precolumn filter. The following gradient conditions were applied. Alanine samples: isocratic elution with 1.9%B, flow rate 0.5ml/min, stop time 17min, column temperature 50°C; lysine system: 0min 0%B, 8min 0%B, 9min 5%B, 11min 5%B, 12min 0%B, flow rate 0.4ml/min, stop time 25min, column temperature 50°C; histidine samples: isocratic elution with 1.7%B, flow rate 0.5ml/min, stop time 30min, column temperature 50°C.

For all systems, detection was operated at 220nm, 4nm bandwidth with a reference wavelength at 550nm, 100nm bandwidth. Each sample was analysed three times.

**Results**

The D- and L- peak areas of the initially racemic alanine, lysine and histidine solutions, reacted sequentially for periods of two hours with fresh vermiculite gel, are presented in Tables 1, 3 and 4 together with D/L ratios and the means and standard deviations of each set of three analyses. Samples of the initial solutions were removed from the first reactant phial in each case, immediately after addition of the solution to the vermiculite gel, and are given as the initial values. All solution D/L ratios were normalised to these analysed starting values and are also reported in the data Tables. The normalised D/L ratios for the amino acids are plotted against the number of exchanges (time) in Fig.1. The solutions of alanine and lysine show a statistically significant decrease in D/L ratios with time during reaction with the vermiculite clay gels, the L-enantiomer being enriched in the supernatant solutions. In contrast, histidine solutions evolve in the opposite sense with D/L ratios increasing as histidine solutions are reacted sequentially with the vermiculite gels. To check this and to investigate the effect of reaction time, a fresh set of racemic histidine solutions was reacted with more fresh gel samples at four hour intervals and the results are also shown in Table 4 and Fig. 1. The 4-hour exchanges show no significant difference from the 2-hour experiments and confirm the increase in D/L ratios with time. The curvature of the slopes reflects the nature of these experiments in which smaller and smaller volumes were decanted onto fresh vermiculite at each stage.

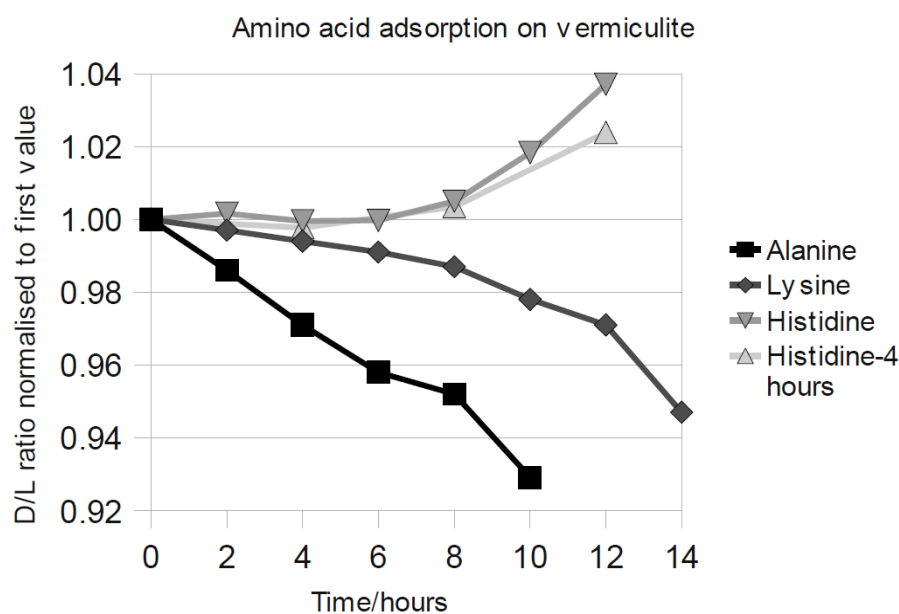

Fig. 1. Changes in D/L ratios of alanine, lysine and histidine solutions reacted sequentially with n-propylNH$_3$Cl vermiculite gel at 32C for periods of 2 hours per reaction together with additional histidine data for repeat reactions using 4 hour intervals.

The difference in behaviour of histidine solutions from alanine and lysine solutions is significant. Since it is known that once seeded with one enantiomer, clay surfaces enhance selectivity for that enantiomer during further crystallization cycles[14] it is thus unlikely that the data presented here are the result of contamination of either the clays, or reagents used, with chiral L-amino acids originating in biological material.

In the light of these results, a further set of experiments was performed to investigate the observed D/L fractionation in more detail. In these, the gels used in the initial experiments were flushed with fresh stock n-propyl ammonium chloride solution at the same concentration. This was carried out for several days, the supernatant solution being replaced every few hours so as to allow the amino acids introduced previously to diffuse out. Samples of the solutions used to flush the gels were taken after each washing and analysed for their amino acid content. No amino acids were observed by HPLC analysis at detection level. These recycled gels were then used for a repeated set of experiments in which amino acid solutions were again added in 2 hour intervals with samples taken and gels changed as before. The results are presented in the same form in Tables 2 and 3. The t=0 normalised D/L are presented in Fig.2 together with the results obtained from the experiments using fresh gels for comparison.

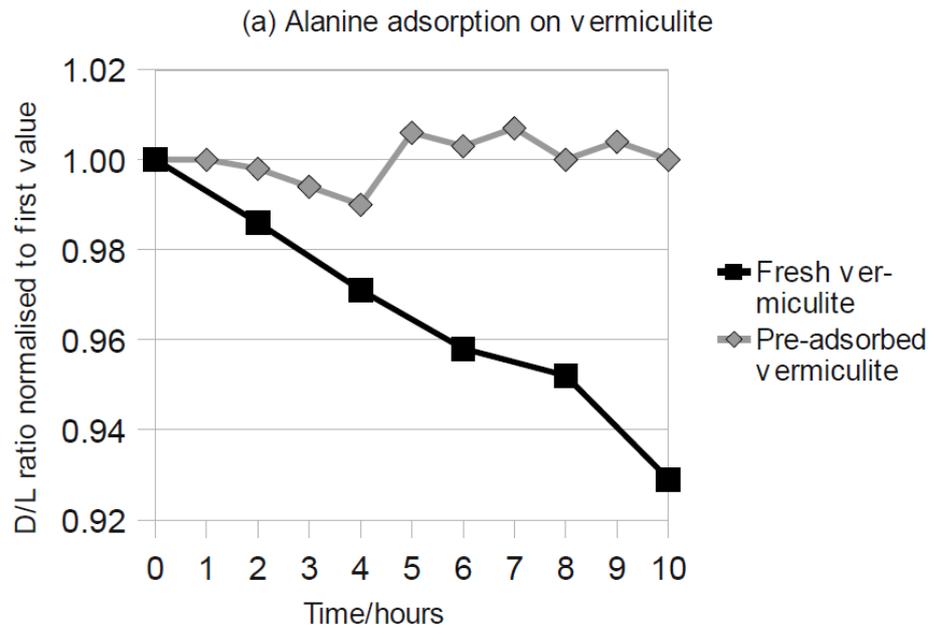

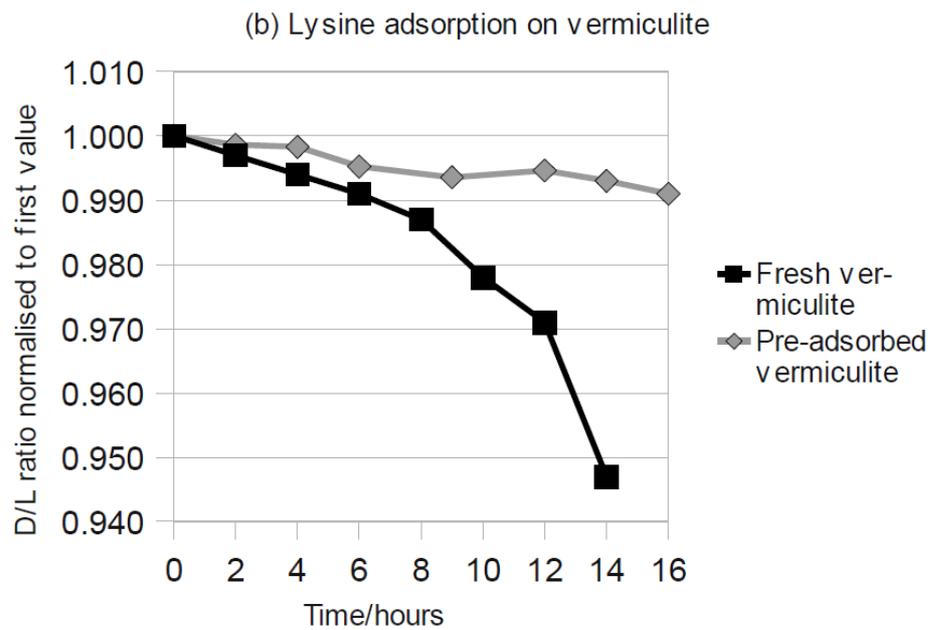

Fig. 2. D/L ratios of alanine and lysine solutions reacted sequentially at 32C for periods of 2 hours per reaction with n-propylNH$_3$Cl vermiculite gel pre-adsorbed with the respective amino acids. Data for fresh gels are given for comparison.

**Discussion**

It is well known that substitution of the inorganic interlayer cations that occur naturally in clays with small organic cations, profoundly alters the clays' adsorption characteristics. For example early work showed that intercalation of tetramethylammonium and tetraethylammonium ions in montmorillonite changes its adsorption properties significantly[52] and in contrast to the original clay, aliphatic and aromatic organic molecules, as well as polar organic species, are readily adsorbed onto the modified interlayer[53]. Short chain alkyl ammonium cations such as n-propyl and n-butyl behave similarly and for highly crystalline materials, 3-D ordered structures can be produced that have interlayer cavities (galleries) with adjoining nanopore surfaces (pillars) consisting of the surface of the organic molecule and the silicate surface of the 2:1 TOT layers[54]. The adsorption characteristics of these pillared clays are determined by the shape, size and, in some cases, the charge of the organic cation[55].

In the case of the n-propyl ammonium gels used in the present study, high resolution neutron diffraction shows that the initial organic cations present are clustered towards the centre of the inter-layer region[47]. In contrast, for the amino acid histidine, Monte-Carlo and molecular dynamics simulations show that a high proportion of the amino acid is adsorbed at the negatively charged TOT siloxane surface. Emission spectra also show that pyrenylbutyl-trimethylammonium clusters on clay surfaces and that it is not adsorbed randomly[56]. Density functional theory simulations have also been used to calculate the adsorption energetics of amino acids on clay surfaces. The results for D- and L-alanine enantiomers of N-formyl alanine amide on the surface of the Fe-bearing clay nontronite, show that binding of the L- form with the clay surface used is 25 kJ/mol more favourable than that of the system with the D-enantiomer. This is also reflected in a smaller $d_{001}$ spacing [57] despite the larger interlayer spacing required by simple geometry of the L-amino acid configuration. This result is of course reversed for a model clay surface of opposite chirality. Recently, changes in d-spacing caused by adding amino acid solutions to n-propyl ammonium vermiculite gels have been measured experimentally by neutron diffraction[48], and again a chiral effect was observed with $d_{001}$ expanding anti-osmotically more for added L-histidine than for added D-histidine.

The results shown in Fig. 1. are in accord with these data. In contrast to alanine and lysine solutions whose D/L ratios decrease as reaction with vermiculite takes place, histidine reacts in the opposite sense. The D/L ratios of histidine solutions increase during reaction with these vermiculite gels. This implies that L-histidine is more strongly adsorbed in vermiculite inter-layers than D-histidine as is shown by neutron scattering data carried out in real time in the vermiculite-solution system.

The contrast between the behaviour of amino acid solutions reacted with pre-adsorbed gels and the results of the experiments performed on fresh, unreacted gels, suggests that amino acids, once adsorbed on the vermiculite gel interlayer, are quite strongly bound and more difficult to displace as is also implied by DFT simulations[57]. For alanine, no D/L fractionation was observed when using previously exposed gels despite the extended length of the observations. The lysine data show a very small decrease in D/L that is just significant at the level of analytical precision. In addition to this significant difference in behaviour, and the implications of the DFT and MD simulations, neutron diffraction data showed no significant reverse shift in the *d*-spacing on going from L- to D- enantiomer over the timescale of the whole experiment (~ 6 hours)[48]. All these data suggest that in the amino acid-clay system, there is a lack of short term reversibility and this lends support to the proposition that amino acids are bound directly and strongly to the clay surface.

The data in Fig. 1 show not only that n-propyl ammonium gels exhibit chiral selectivity, but that the effect differs according to the amino acid. Pure crystalline vermiculite is monoclinic with the space group c2/m and so should not on its own exhibit chiral selectivity. Moreover, all experiments designed to investigate chiral fractionation are subject to the possible effects of contamination by natural biological materials. The difference in behaviour of histidine which shows preferential adsorption of the L-enantiomer and alanine and lysine, for both of which the D-enantiomer adsorbs more readily, suggests that in these experiments, contamination with biological material did not play a part in determining the chiral selectivity observed.

In the context of biochirality, enantiomeric excesses can only result if chiral selection operates on the interlayer itself. Chiral modification of solid surfaces by symmetry removal through adsorption is well known e.g.[58-59] and surfaces can play an important role in molecular self-organization[60-61]. It is therefore not surprising that clay interlayer surfaces exhibit chiral activity as shown by the present results. Molecular organization in thin films is of considerable interest in nanochemistry research. Thin films on clays can be produced by the Langmuir-Blodgett method at the air-solution interface when volatile solvents are used or by Layer-by-Layer (L-b-L) assembly on clay microplatelets in suspension[62]. In the present case, both the n-propyl $NH_3$ cations and added amino acids provide organic molecules capable of molecular organization. Polarized UV-visible spectroscopy of clay-alkylammonium-dye multilayers shows that the dye molecules present are preferentially adsorbed on the clay inter-layer and show molecular organization that is non-centrosymmetric[63]. We believe that the organization of amphiphilic organic molecules on clay platelets to form ordered nanofilms that carry important functional properties such as chirality and optical non-linearity is key to understanding how clays may act in the natural environment to induce chiral selectivity. In the present case, D/L depletion was observed for alanine and lysine solutions and D/L enrichment for histidine solutions. We note that the specific optical rotations of the L-enantiomers (all c= 2.0 in $H_2O$) of L-Ala ( [α] = +1.8[25]) and L-Lys ( [α] = +13.5[25]) on the one hand and L-His ( [α] = -38.5[25]) on the other, are of opposite sign[64]. If H-bonding of the amino acids in solution or near the acidic siloxane interlayer surface leads to dimer or oligomer formation, steric factors could lead also to changes in the specific optical activity. This might play a significant role in the adsorption as has been oberved for aniline derivatives in montmorillonite layers[65]. For the sensitive vermiculite model clay used here, repeated exposure of initially racemic solutions to the clay produce both D- and L-enrichment in subsequent solutions. The strength of the D/L fractionation may in part be related to the hydrophobicity of the different amino acids and how this forms a part of the energetics of the host-guest-guest interactions in the nanolayer. The small grain size and poor crystallographic order of many clays makes the study of intercalation behaviour very difficult and how other clay minerals may behave, given sufficient time to react, is unknown. In a simulated hydrothermal environment in which fluid circulation took place repeatedly between hot and cold regions, differences in D- and L-alanine racemisation were observed and again L-alanine was slightly enriched in solution compared with D-alanine[66]. The concentrations required for this process can be achieved by adsorption in clay interlayers. Both in the lagoonal clay-brine scenario in the primitive ocean postulated here and in hydrothermal environments, repeated exposure to mineral surfaces selectively leads to enantiomeric excess in favour of L-alanine[66].

Considerable interest is currently being shown in the properties of ordered molecular systems in the solid state. Stacking sequences in kaolinite have been identified by electron back-scattered diffraction [67] and there is interest in the chiral recognition properties of clays[68]. The key to understanding the ability of clays to undertake chiral recognition lies in their properties as rigid 2-D silicate layers separated by nanopores. Molecular orientation and self-organization in the interlayer are important via guest-guest interactions among the intercalated organic species. Atomistic simulations show that while organic complexes are strongly adsorbed in the first molecular layer, the formation of double-layers is important in clays and that such double layers exhibit ordering and select molecules of the same chirality[69]. This is in accord with our observations of the chiral selectivity of vermiculite gels.

**Conclusions**

Amino acids adsorb stably on clay surfaces present in nature. The data presented here show that vermiculite clay gels with large interlayer spacings act within hours as chiral amplifiers that sequentially change the D/L ratios of solutions containing amino acids. The kinetics of adsorption and desorption appear to be different as do the behaviours of alanine, lysine and histidine. Overall chiral selectivity requires a net chiral symmetry operator. Although one such possibility would be chiral energetic differences caused by magnetochiral dichroic effects of the Earth's magnetic dipole on Fe-containing minerals, such effects are very small [70]. We propose instead that chiral selection occurs by means of non-centrosymmetric clay film nanolayers in the inter-layer region. This could take place naturally in a clay-rich lagoonal environment containing simple prebiotic molecules at ambient planetary pressures and temperatures. With repetition of these D/L fractionations over geological periods of time in hydrous clay-rich conditions under which exchanged fluids could migrate, the chiral amplification could be large. Seeding of clay/solution nanofilm multilayers with amino acids having a chiral excess whether from space[25, 28] or produced in a modified Miller-Urey synthesis[71] in the presence, for example, of polarised light or dust particles with chiral centres, could readily lead to further enhancement of the chiral properties of clay-organic nanofilms[72-73]. Such activated clay films have the capacity not only to concentrate amino acids from dilute solution, but to induce very significant chiral separation of amino acids in solutions in contact with the clay particles when solutions undergo repeated cycling. This may be a reason for the chiral selectivity observed in iso-valine in the Murchison and Orgueil meteorites that is related to the presence of hydrous minerals[25]. Considering these observations together with the role of clays in catalysing the oligomerization of activated RNA monomers[34] and the need to organize amphiphilic molecules and lipids in the development of primitive cells[74], processes such as these, in which molecular ordering occurs to yield natural clay inter-layer nanofilms that lead to rapid development of chiral excesses in amino acids, may have played some role in the development of the chiral amplification that occurred during selection for biochirality.

**Acknowledgements**
This work was financially supported by the Austrian Federal Ministry for Science and Research (BMWF, Grant Nr. 45.530/0003-11/6a/2007) and by the Austrian Science Foundation (Fonds zur Förderung der wissenschaftlichen Forschung, Projekt P19334-N17) which is gratefully acknowledged. We are very grateful to Dr N. Skipper for making available samples of n-propylNH$_3$Cl-vermiculite. DGF gratefully acknowledges financial support from Worcester College, Oxford, warmly thanks Neil Skipper, Martin Smalley and Chris Greenwell for many hours of nocturnal discussions around neutron beam lines and Professor Rode for his generous support and invitation to work in Innsbruck.

Table 1: Results of alanine measurements using fresh vermiculite with exchange of gels every 2 hours.

| Time [hours] | Peak area L-Ala | Peak area D-Ala | D/L Ratio | Average | Std. Dev. | Normalised Average |
|---|---|---|---|---|---|---|
| 0 | 9451.0 | 10296.2 | 1.089 | | | |
| | 9429.9 | 10365.6 | 1.099 | 1.090 | 0.009 | 1.000 |
| | 9480.4 | 10252.1 | 1.081 | | | |
| 2 | 9870.7 | 10768.0 | 1.091 | | | |
| | 9325.6 | 9923.1 | 1.064 | 1.075 | 0.014 | 0.986 |
| | 9230.3 | 9868.9 | 1.069 | | | |
| 4 | 8949.2 | 9430.5 | 1.054 | | | |
| | 8911.9 | 9492.6 | 1.065 | 1.059 | 0.006 | 0.971 |
| | 8859.1 | 9362.9 | 1.057 | | | |
| 6 | 8739.4 | 9131.6 | 1.045 | | | |
| | 8538.1 | 8988.7 | 1.053 | 1.045 | 0.008 | 0.958 |
| | 8506.6 | 8813.4 | 1.036 | | | |
| 8 | 8203.2 | 8490.5 | 1.035 | | | |
| | 8112.5 | 8453.8 | 1.042 | 1.038 | 0.004 | 0.952 |
| | 8061.0 | 8363.8 | 1.038 | | | |
| 10 | 7495.5 | 7606.5 | 1.015 | | | |
| | 7501.2 | 7589.0 | 1.012 | 1.013 | 0.002 | 0.929 |
| | 7465.3 | 7552.1 | 1.012 | | | |

Table 2: Results of alanine measurements using pre-adsorbed vermiculite with exchange of gels every hour.

| Time [hours] | Peak area L-Ala | Peak area D-Ala | D/L Ratio | Average | Std. Dev. | Normalised Average |
|---|---|---|---|---|---|---|
|   | 10813.7 | 10570.7 | 0.978 |       |       |       |
| 0 | 10422.0 | 10280.0 | 0.986 | 0.985 | 0.007 | 1.000 |
|   | 10422.8 | 10340.5 | 0.992 |       |       |       |
|   | 9485.6  | 9347.8  | 0.985 |       |       |       |
| 1 | 9262.9  | 9105.5  | 0.983 | 0.985 | 0.002 | 1.000 |
|   | 9205.4  | 9082.4  | 0.987 |       |       |       |
|   | 9250.9  | 9153.2  | 0.989 |       |       |       |
| 2 | 9303.3  | 9129.6  | 0.981 | 0.984 | 0.005 | 0.998 |
|   | 9263.4  | 9078.7  | 0.980 |       |       |       |
|   | 8707.9  | 8477.6  | 0.974 |       |       |       |
| 3 | 8728.9  | 8586.8  | 0.984 | 0.979 | 0.005 | 0.994 |
|   | 8759.4  | 8589.5  | 0.981 |       |       |       |
|   | 8834.8  | 8682.1  | 0.983 |       |       |       |
| 4 | 8878.4  | 8679.3  | 0.978 | 0.975 | 0.009 | 0.990 |
|   | 9487.7  | 9164.9  | 0.966 |       |       |       |
|   | 8989.1  | 8771.3  | 0.976 |       |       |       |
| 5 | 8729.6  | 8710.1  | 0.998 | 0.991 | 0.013 | 1.006 |
|   | 8749.0  | 8739.5  | 0.999 |       |       |       |
|   | 8911.9  | 8735.1  | 0.980 |       |       |       |
| 6 | 8911.9  | 8888.3  | 0.997 | 0.988 | 0.009 | 1.003 |
|   | 8885.3  | 8761.1  | 0.986 |       |       |       |
|   | 8865.8  | 8933.2  | 1.008 |       |       |       |
| 7 | 9011.7  | 8867.9  | 0.984 | 0.992 | 0.013 | 1.007 |
|   | 8958.6  | 8825.5  | 0.985 |       |       |       |
|   | 8599.7  | 8443.3  | 0.982 |       |       |       |
| 8 | 8786.9  | 8681.5  | 0.988 | 0.985 | 0.003 | 1.000 |
|   | 8814.8  | 8683.3  | 0.985 |       |       |       |
|   | 8907.2  | 8903.2  | 1.000 |       |       |       |
| 9 | 8982.4  | 8842.4  | 0.984 | 0.989 | 0.009 | 1.004 |
|   | 9005.0  | 8862.8  | 0.984 |       |       |       |
|   | 9244.3  | 9129.7  | 0.988 |       |       |       |
| 10| 9209.7  | 9049.0  | 0.983 | 0.986 | 0.003 | 1.000 |
|   | 9232.7  | 9111.0  | 0.987 |       |       |       |

Table 3: Results of lysine measurements using fresh (upper part of table) and pre-adsorbed (lower part) vermiculite with exchange of gels every 2 hours (only one measurement each for the experiments with pre-adsorbed vermiculite).

Lysine measurements using fresh vermiculite

| Time [hours] | Peak area L-Lys | Peak area D-Lys | D/L Ratio | Average | Std. Dev. | Normalised Average |
|---|---|---|---|---|---|---|
| 0 | 12134.4 | 12873.7 | 1.061 | | | |
|   | 12237.8 | 12949.2 | 1.058 | 1.059 | 0.002 | 1.000 |
|   | 12315.9 | 13032.5 | 1.058 | | | |
| 2 | 10498.9 | 11101.2 | 1.057 | | | |
|   | 11683.1 | 12360.9 | 1.058 | 1.056 | 0.003 | 0.997 |
|   | 11742.5 | 12364.4 | 1.053 | | | |
| 4 | 11012.7 | 11608.9 | 1.054 | | | |
|   | 11043.7 | 11605.4 | 1.051 | 1.053 | 0.002 | 0.994 |
|   | 11120.5 | 11706.1 | 1.053 | | | |
| 6 | 9975.5 | 10451.5 | 1.048 | | | |
|   | 10270.2 | 10801.6 | 1.052 | 1.050 | 0.002 | 0.991 |
|   | 10429.2 | 10953.4 | 1.050 | | | |
| 8 | 9378.5 | 9811.8 | 1.046 | | | |
|   | 9613.3 | 9996.3 | 1.040 | 1.045 | 0.005 | 0.987 |
|   | 9752.1 | 10235.2 | 1.050 | | | |
| 10 | 8554.1 | 8905.6 | 1.041 | | | |
|   | 8798.2 | 9109.9 | 1.035 | 1.035 | 0.006 | 0.978 |
|   | 8771.4 | 9030.8 | 1.030 | | | |
| 12 | 6935.8 | 7131.2 | 1.028 | | | |
|   | 7227.3 | 7459.1 | 1.032 | 1.028 | 0.004 | 0.971 |
|   | 7248.4 | 7424.7 | 1.024 | | | |
| 14 | 4455.0 | 4460.0 | 1.001 | | | |
|   | 4487.1 | 4499.6 | 1.003 | 1.003 | 0.003 | 0.947 |
|   | 4514.7 | 4542.1 | 1.006 | | | |

Lysine measurements using pre-adsorbed vermiculite

| Time [hours] | Peak area L-Lys | Peak area D-Lys | D/L Ratio | Average | Std. Dev. | Normalised Average |
|---|---|---|---|---|---|---|
| 0 | 16844.2 | 17884.7 | 1.062 | 1.062 | - | 1.000 |
| 2 | 15854.6 | 16811.0 | 1.060 | 1.060 | - | 0.999 |
| 4 | 14568.4 | 15441.8 | 1.060 | 1.060 | - | 0.998 |
| 6 | 13180.1 | 13927.5 | 1.057 | 1.057 | - | 0.995 |
| 9 | 11564.5 | 12199.8 | 1.055 | 1.055 | - | 0.994 |
| 12 | 10554.0 | 11145.7 | 1.056 | 1.056 | - | 0.995 |
| 14 | 8202.1 | 8647.8 | 1.054 | 1.054 | - | 0.993 |
| 16 | 5989.8 | 6302.6 | 1.052 | 1.052 | - | 0.991 |

Table 4: Results of histidine measurements using fresh vermiculite with exchange of gels every 2 hours (upper part of table) and every 4 hours (lower part).

Histidine measurements using fresh vermiculite with exchange of gels every 2 hours

| Time [hours] | Peak area L-His | Peak area D-His | D/L Ratio | Average | Std. Dev. | Normalised Average |
|---|---|---|---|---|---|---|
|   | 21762.7 | 21255.7 | 0.977 |   |   |   |
| 0 | 21757.3 | 21196.8 | 0.974 | 0.972 | 0.007 | 1.000 |
|   | 21969.3 | 21184.6 | 0.964 |   |   |   |
|   | 20990.3 | 20355.0 | 0.970 |   |   |   |
| 2 | 21253.3 | 20939.5 | 0.985 | 0.973 | 0.011 | 1.002 |
|   | 20993.2 | 20260.2 | 0.965 |   |   |   |
|   | 20447.4 | 19345.1 | 0.946 |   |   |   |
| 4 | 19753.1 | 19452.2 | 0.985 | 0.971 | 0.022 | 1.000 |
|   | 19792.6 | 19458.1 | 0.983 |   |   |   |
|   | 19557.1 | 18924.3 | 0.968 |   |   |   |
| 6 | 19410.8 | 18754.3 | 0.966 | 0.971 | 0.008 | 1.000 |
|   | 19138.8 | 18767.0 | 0.981 |   |   |   |
|   | 17925.0 | 17669.9 | 0.986 |   |   |   |
| 8 | 18344.2 | 17650.4 | 0.962 | 0.977 | 0.013 | 1.005 |
|   | 18026.8 | 17698.7 | 0.982 |   |   |   |
|   | 16766.0 | 16643.8 | 0.993 |   |   |   |
| 10 | 16682.6 | 16604.9 | 0.995 | 0.990 | 0.008 | 1.018 |
|   | 17027.0 | 16696.0 | 0.981 |   |   |   |
|   | 14317.0 | 14678.4 | 1.025 |   |   |   |
| 12 | 14155.9 | 14208.2 | 1.004 | 1.008 | 0.016 | 1.037 |
|   | 14552.7 | 14472.4 | 0.994 |   |   |   |

Histidine measurements using fresh vermiculite with exchange of gels every 4 hours

| Time [hours] | Peak area L-His | Peak area D-His | D/L Ratio | Average | Std. Dev. | Normalised Average |
|---|---|---|---|---|---|---|
|   | 25385.7 | 23876.2 | 0.941 |   |   |   |
| 0 | 22637.3 | 21949.1 | 0.970 | 0.957 | 0.015 | 1.000 |
|   | 21955.8 | 21071.5 | 0.960 |   |   |   |
|   | 21183.6 | 20308.9 | 0.959 |   |   |   |
| 4 | 21515.2 | 20074.4 | 0.933 | 0.954 | 0.020 | 0.998 |
|   | 21200.0 | 20597.9 | 0.972 |   |   |   |
|   | 19305.8 | 19121.5 | 0.990 |   |   |   |
| 8 | 20926.6 | 20124.9 | 0.962 | 0.960 | 0.031 | 1.003 |
|   | 20530.6 | 19045.3 | 0.928 |   |   |   |
|   | 18152.2 | 17691.9 | 0.975 |   |   |   |
| 12 | 18747.4 | 18490.9 | 0.986 | 0.979 | 0.006 | 1.024 |
|   | 19017.0 | 18589.4 | 0.978 |   |   |   |